# Floods impact dynamics quantified from big data sources

David Pastor-Escuredo[1,2,*], Yolanda Torres[2], María Martínez[2],
Pedro J. Zufiria[2]

[1] *LifeD Lab*
[2] *Universidad Politécnica de Madrid*
*corresponding author: david@lifedlab.org*

Natural disasters affect hundreds of millions of people worldwide every year. Early warning, humanitarian response and recovery mechanisms can be improved by using big data sources. Measuring the different dimensions of the impact of natural disasters is critical for designing policies and building up resilience. Detailed quantification of the movement and behaviours of affected populations requires the use of high granularity data that entails privacy risks. Leveraging all this data is costly and has to be done ensuring privacy and security of large amounts of data. Proxies based on social media and data aggregates would streamline this process by providing evidences and narrowing requirements. We propose a framework that integrates environmental data, social media, remote sensing, digital topography and mobile phone data to understand different types of floods and how data can provide insights useful for managing humanitarian action and recovery plans. Thus, data is dynamically requested upon data-based indicators forming a multi-granularity and multi-access data pipeline. We present a composed study of three cases to show potential variability in the natures of floodings,as well as the impact and applicability of data sources. Critical heterogeneity of the available data in the different cases has to be addressed in order to design systematic approaches based on data. The proposed framework  establishes the foundation to relate the physical  and socio-economical impacts of floods.

**INTRODUCTION**

Natural disasters such as floods, storms or earthquakes affect hundreds of millions of people worldwide every year. Humanitarian action can be potentially improved with dynamic in-situ information across the disaster timeline supporting the different stages of disaster management cycle: preparedness, warning, response, evaluation and mitigation. Inserting data-driven mechanisms throughout the management of disasters could improve decision making, monitoring and evaluation activities. A deeper understanding of the behavior at different scales for different disasters is a necessary step to build up more resilient communities and societies. Recently, the use of Big Data has allowed to quantify behaviors and patterns caused by disasters. However, most of these studies are still no real-time. A comprehensive framework considering both the necessary data-insight pipelines and the access-privacy limitations is required to move forwards to the implementation of real-time systems.

The exponential increase in the penetration of mobile phones and the use of social media has motivated the analysis of this big data sources to better understand natural disasters, so that new opportunities for obtaining such indicators have emerged [1]. Anonymized and aggregated Call Detail Records -CDRs- have enabled large scale analysis of social patterns, specially studies on human mobility [2-3]. Humanitarian applications based on this big data source cover, for instance, the monitoring of disease outbreaks [4] or the quantification of social patterns and mobility during disasters [5-8]. Satellite imagery has been used extensively and, in the recent years, several studies have exploited this resource to estimate poverty proxies [9-10]. Call Detail Records have been also used to estimate socio-economic profiles or mobility profiles related to livelihoods [11-13]. The public availability of social media data has allowed its beneficial use for social purposes such as unemployment estimations [14] or assessing the damage during natural disasters [15].

In this work, we address how to quantify the impact of floods dynamically considering the issue of accessing private sector data in real-time. In the methods section, data sources and technical descriptions of the methods are presented. The results section shows the application of the methods and data to the study cases selected. Discussion presents an integrative argumentation of the results, limitations and scope of this work.

**METHODS**

**Flood segmentation**

Flood segmentation consisted of two steps, first an area extraction based on satellite images and then a depth estimation using Digital Elevation Models.

*Satellite image analysis for flooded area extraction:*

The satellite images were downloaded from Planet by drawing a polygon as area of interest. The Analytic PLANETSCOPE products provided by PLANET are 4-band (RGB, NIR), high resolution images, with ground sample distance (GSD) of 3m [16]. This, along with the rather low cloud cover rate (0%-25%), allowed for direct photointerpretation. For the flooded area estimation, we selected two images, before and after the floods, in order to compare the pre- and post-event situations.

The images were analyzed within a Geographical Information System (GIS) [17]. Two band combinations were interchanged, as needed for photointerpretation: true color (Red=band 3; Green=band 2; Blue=band 1) and infrared (Red=band 4; Green=band 3; Blue=band 2). The infrared combination highlights water bodies in light blue, making it easier to identify flooded areas. Water bodies have been digitized in pre- and post-event images, to create two hydrography polygon layers. Then, the flooded area was obtained by overlaying both polygons, so that its surface could be measured.

*Digital Elevation Model (DEM) integration:*

We gathered and used a SRTM (Shuttle Radar Topography Mission) 30-meter resolution raster with elevation data to the GIS, with the aim of analyzing the effect of the topography on the flood. The raster was downloaded from the USGS Earth Explorer repository.

The hydrography layers were projected over the DEM, in order to convert 2D into 3D features. Then, the Z components were extracted to obtain the mean, maximum and minimum altitude of each river course and flooded area. Again, by comparing pre- and post-event situations, we could evaluate the river rise and water bodies levels.

**Call Detail Records analysis and visualization**

Presence data for France during 2014 was made available by Orange. Presence data was computed by anonymizing and aggregating Call Detail Records -CDRs- [18]. More specifically, presence data consists of the number of people registered at each geolocation (derived from antennas position but modifying the exact position to avoid the allocation of the antennas) and temporal point (with a resolution aggregated from the available timestamp of the CDRs). Similarly, communication volume data was available as the number of calls instead of the number of registered people. We applied this data to study the dynamics of floods impact. For this purpose, we built a multi-resolution dynamic description of people concentration at the *antenna level* with three different temporal resolutions:

*Daily resolution:* This resolution was suitable to generally assess the dynamic population changes due to mobility flows regarding the disaster. It was used as a dynamic census to normalize social media proxies introducing a geo-temporal correction factor. As a dynamic census, it could be used to estimate the number of affected population as an alternative to more static demographics resources.

The daily value of the presence data was computed in two different ways: by averaging the presence within a day or by selecting an interval of hours (20h - 23h) to cumulate the presence data. The computation based on an interval of interest was finally used in this work.

*Weekly resolution*: This data was suitable for assessing large-scale flows of people going in and out the region flooded and the surrounding cities, removing the effects of weekly mobility patterns. Monthly resolution could be also applied for similar purposes. The average of daily presence is used to build the weekly resolution.

*Hourly resolution*: Once the flood is detected, a focus on the 3 prior and posterior days with hourly resolution enabled the visualization of population movements during the disaster. Night hours (1 am - 8 am) were discarded since they introduce a sharp gradient in the computation of the statistics preventing for a correct visualization. The original resolution of the data provided was hourly.

The multi-resolution descriptions from presence data were visualized using z-scores for each geolocation (antenna level) along the different temporal series generating both temporal and static maps with GIS.

**Social media semantic analysis**

Data provided by Crimson Hexagon -CH- was used to make social media analysis during floods [19]. We used buzz monitors from the "ForSight" tool, retrieving posts from all available social media platforms to extract three different types of proxies:

*Awareness proxy*: This signal was obtained by filtering posts by geolocation and using keywords: {"flood", "weather", "rain", "water", "river"}. The translation into Spanish and French was used for the floods occurred in Mocoa and Montpellier respectively.

*Total posts*: This signal was obtained by filtering posts by geolocation using a bounding box and longitude and latitude parameters as keywords in the search.

*Damage proxy*: This signal was obtained by filtering posts by geolocation and keywords: {"insurance", "property", "damage"}. The translation into Spanish and French was used for the floods occurred in Mocoa and Montpellier respectively.

The social media proxies can be calibrated to have a first estimation of affected population with a normalization term that contains the dynamic census extracted from CDR aggregates as shown in Fig. 1.

$$\text{Normalized Awareness} = \frac{\text{Posts Flooding}}{\text{Total Posts}} \times \frac{\text{Social media users}}{\text{census or CDRcensus}}$$

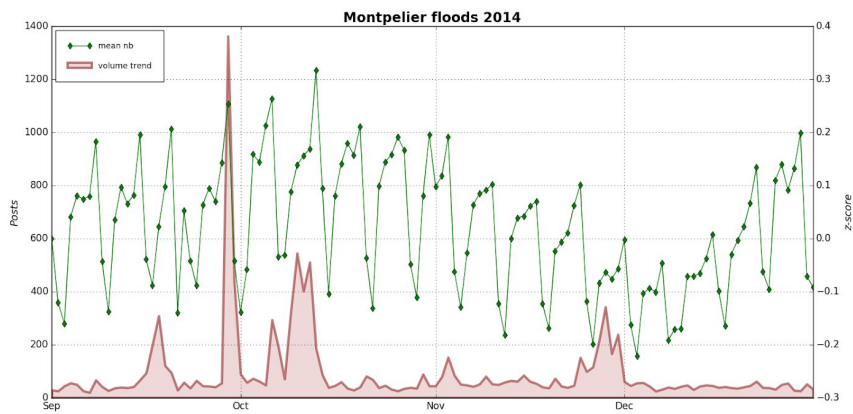

*Figure 1: red: time series of total volume of posts containing the keywords used to calculate the awareness proxy related (before normalization). green: CDR time series of average number of people per antenna at a daily resolution in a field of view that covers the affected area that was segmented with an online interactive tool and exported as geojson.*

Sentiment and emotion analysis was performed using the readily-available modules of the "ForSight" tool. Due to filtering steps performed by CH, the exact geolocation of the posts was not available for this work.

**Social media network analysis**

Bulk download of the posts allowed to further analyze the network built from the posts in twitter, although limited to a random sample of 10,000 post for each download. Although posts from some social media platforms can be openly accessed, the historic database of CH allowed to have posts from all the cases.

Nodes representing the users making posts were connected through edges with nodes representing retweeted or mentioned addresses. We built a time-evolving network throughout the disaster to analyze the dynamics of the links established by the posts. Through time, a timeline of links for each user was created, differentiating both types of nodes. We also annotated and differentiated each node by the gender of the user. We then applied k-means clustering method to profile the dynamics of the two types of nodes obtaining different classes of nodes according to their linking timeline. Aggregating the links of each profile returned different social media patterns appearing as a response to the floods.

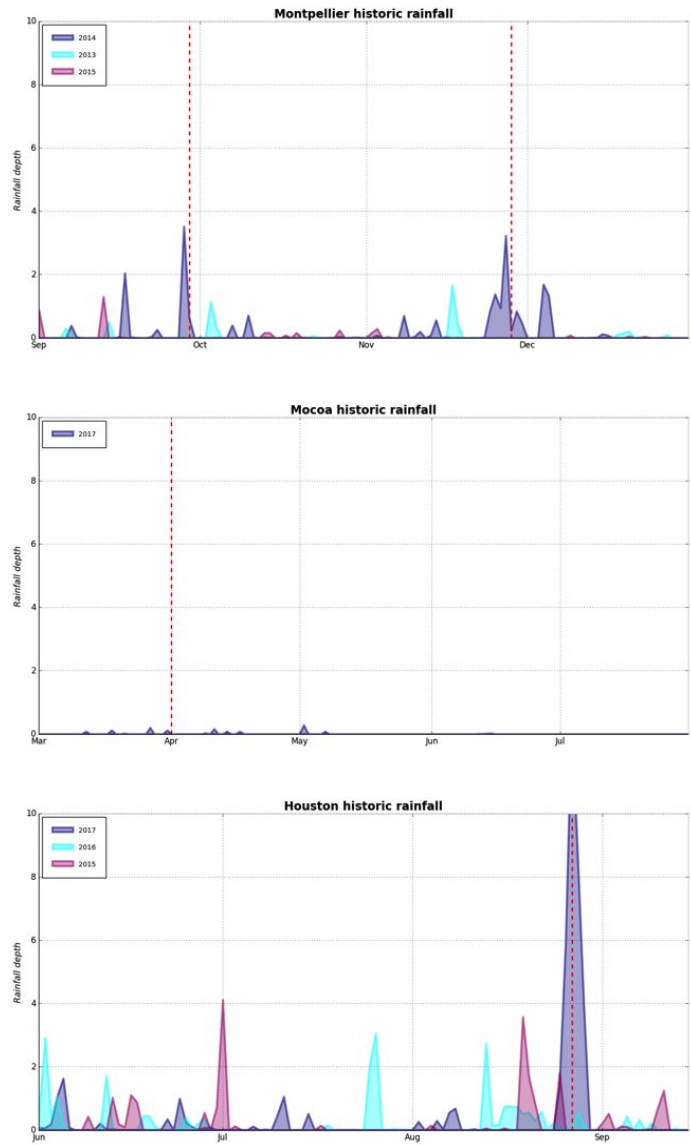

*Figure 2: Historic profile of rainfalls (SE data) in the three study cases selected. Dark blue indicates the year of the flood and the vertical dashed lines the day of the flood as retrieved from news.*

**Rainfall measurements in floods**

Historic profile of rainfalls (Fig. 2) in the affected areas were built using Schneider Electric -SE- web service. The service selects the closest station and returns the data for web queries provided the time interval and the geolocation of the approximated center of the affected area. Earthwork network data is also suitable to obtain rainfall measurements but it was not included in this work so far.

The selected cases were determined by the analysis of news and exploration of data sources, considering different types of causes for flooding.

**RESULTS**

**Framework to interconnect data sources throughout disaster management**

Measuring the impact of floods implies tackling the problem of real-time access of privacy-risky data during disasters. We designed a framework with a multi-granularity and multi-access design based on evidences to retrieve baseline information and request access to high granularity data.

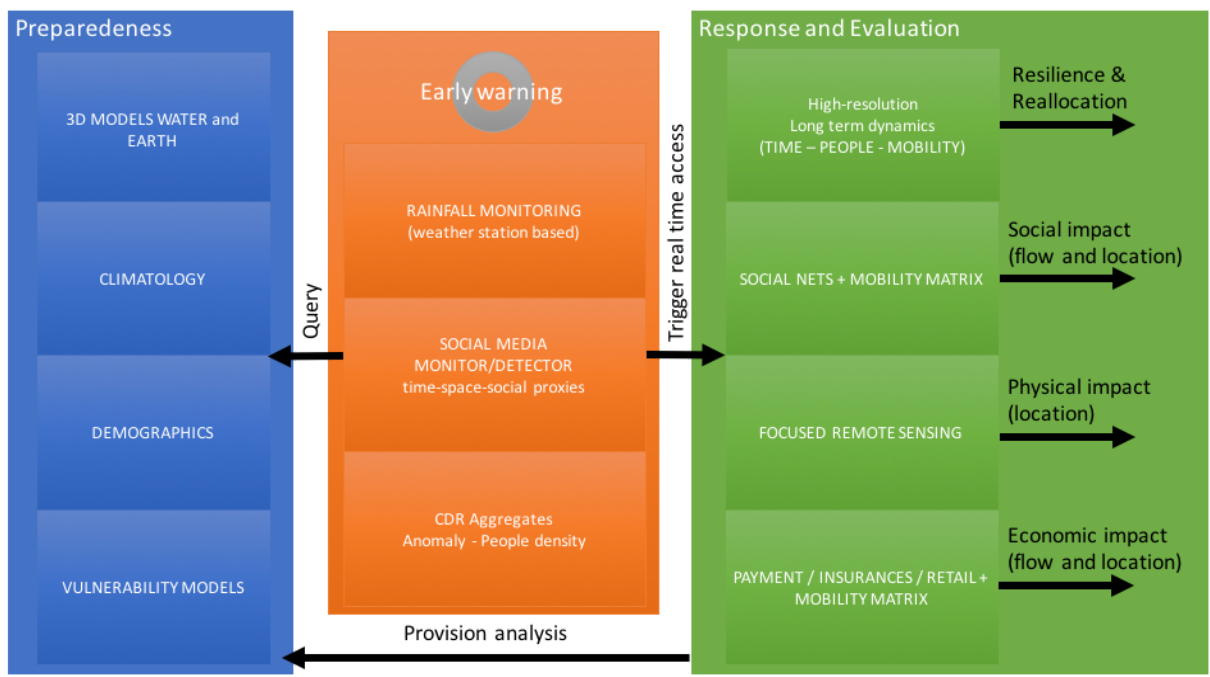

*Figure 3: Framework mapping stages in the disaster management and the necessary data sources. In this framework, the detection module triggers the requests to readily-available data and provides evidences to request high-granularity data.*

As shown in Figure 3, the iteratively running detection and warning module, implemented with open social media data and data aggregates to be available in real-time, would trigger other modules. The warning module provided temporal, spatial and social proxies of impact. The temporal proxy determined the timing of the awareness of the population of a flood. Spatial proxies computed from the geolocation of posts would provide an initial mask to calculate physical impact. The social proxy provided estimation of how many people could be affected and basic demographic information for that population.

The preparedness module was designed to provide different baselines (physical, environmental and socio-economic) for which models could be computed. Most part of the data from these modules are currently available but we observed the need to have standards and connected pipelines. The response and evaluation module comprised different submodules taken as input proxies from the detection module and baselines from the preparedness module. We differentiated the impact estimation in four subgroups: reallocation and dynamics, physical impact, social impact and economical impact.

The physical impact submodule aims at characterizing the location affected using resources available. The reallocation and dynamics submodule aims at having a dynamic characterization through time of the social nets to measure recovery periods of the population affected. For this module, high-granularity CDRs was considered to be a valuable data source. This module also feeds the social and economic submodules where the mobility is necessary to properly characterize the impacts of displaced population due to the disaster along time, as a flow-based representation of the affected population. The social and economic submodules can benefit from social media proxies but eventually require high-granularity data sources like CDRs, financial and transactions data or retail data.

**Detection and warning of different types of floods**

We compared the rainfalls estimation extracted from the SE service with the awareness proxy computed with social media data extracted from CH. Peaks in the rainfalls estimation and the awareness proxy synchronized during floods due to torrential rainfalls: Montpellier -France- in September and November 2014 and Houston -USA- in August 2017. However, peaks in the awareness proxy were present during floods due to overflows: Mocoa -Colombia- April 2017, where no rainfalls measurement could provide a timely detection (Fig. 4). The awareness proxy of the floods in Montpellier was enhanced by the normalization with the dynamic census computed from presence data at a daily resolution derived from CDRs (Methods). As shown in Fig. 1, the population significantly changed between the two floods, so the dynamic census allowed for a clear detection by the awareness proxy.

**Physical impact of floods**

Satellite imagery and a Digital Elevation Model were used to compute a 3D characterization of the flood occurred in Mocoa, both surface and elevation at different points on the flood (Methods). These measurements provide quantitative evidence of the physical impact of the flood.

River and flood segmentations were computed from Planet image data available from March 14th (pre-event) and April 10th (post-event) while the flood happened on April 1st according to news and the awareness proxy (Fig. 5). This delay was considered critical for a timely and accurate 3D characterization of the flood at its highest impact. The geospatial resolution of the DEM (30 m) and the lack of standards across sources for elevation values prevented for a full 3D reconstruction of the flood. Data for the other cases was analyzed but the difficulties in the characterization of water in urban areas or the presence of dense clouds did not allow to reproduce the same results. The  were overlaid with spatial descriptions of the region (Fig. 5), available as raster data [16, 20], to have a better understanding of the impact in infrastructure and demographics.

Although physical impact quantification was precisely achieved using imagery and geospatial resources, social media proxies could help focusing efforts to reduce the temporal gap in acquisition.

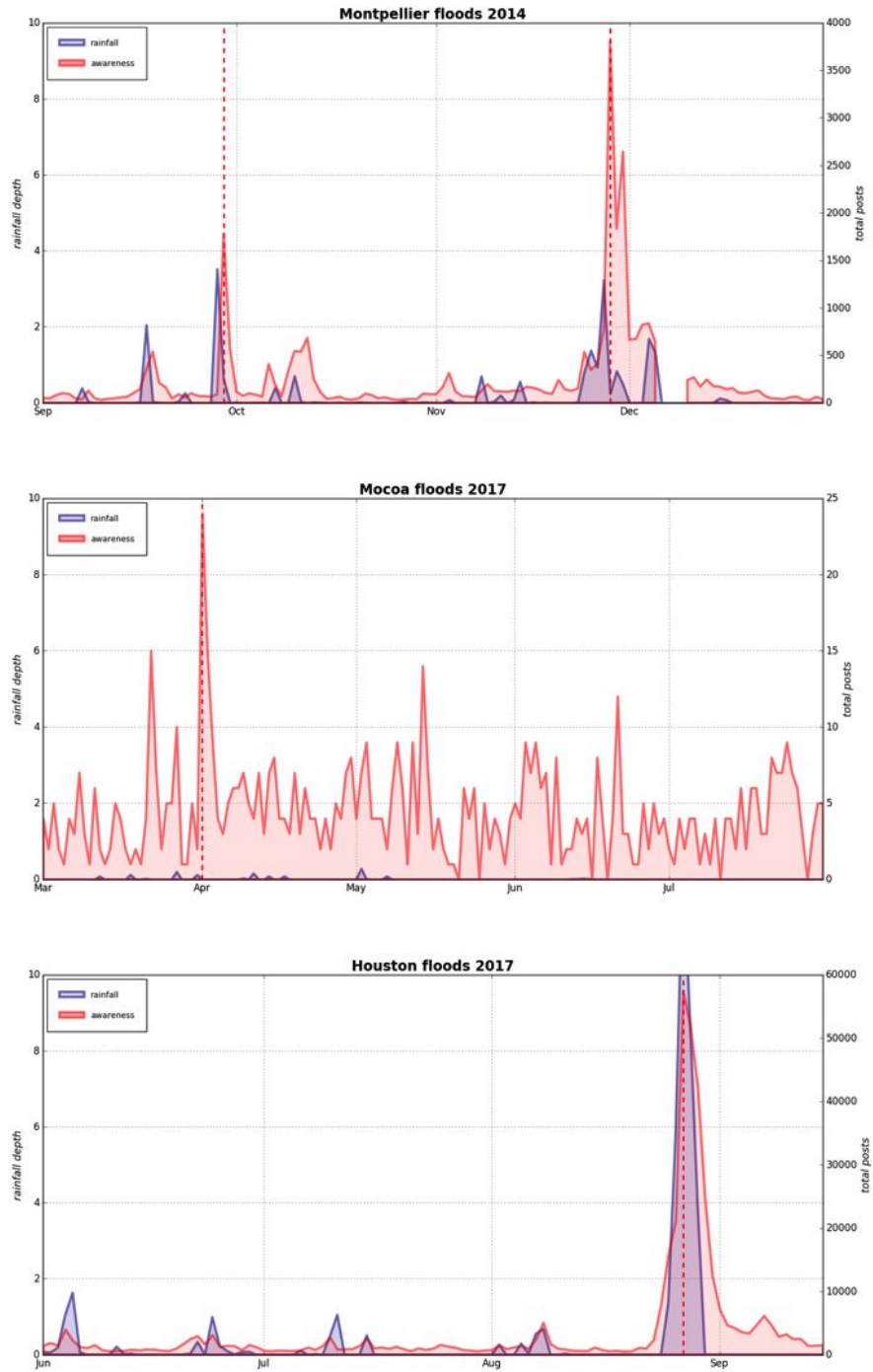

*Figure 4: Awareness proxy (red) over Rainfall measurement (blue). The awareness proxy for the Montpellier flood was normalized by CDR aggregate as indicated in the Methods section and Figure 1. Vertical dashed lines indicate the occurrence of a flood*

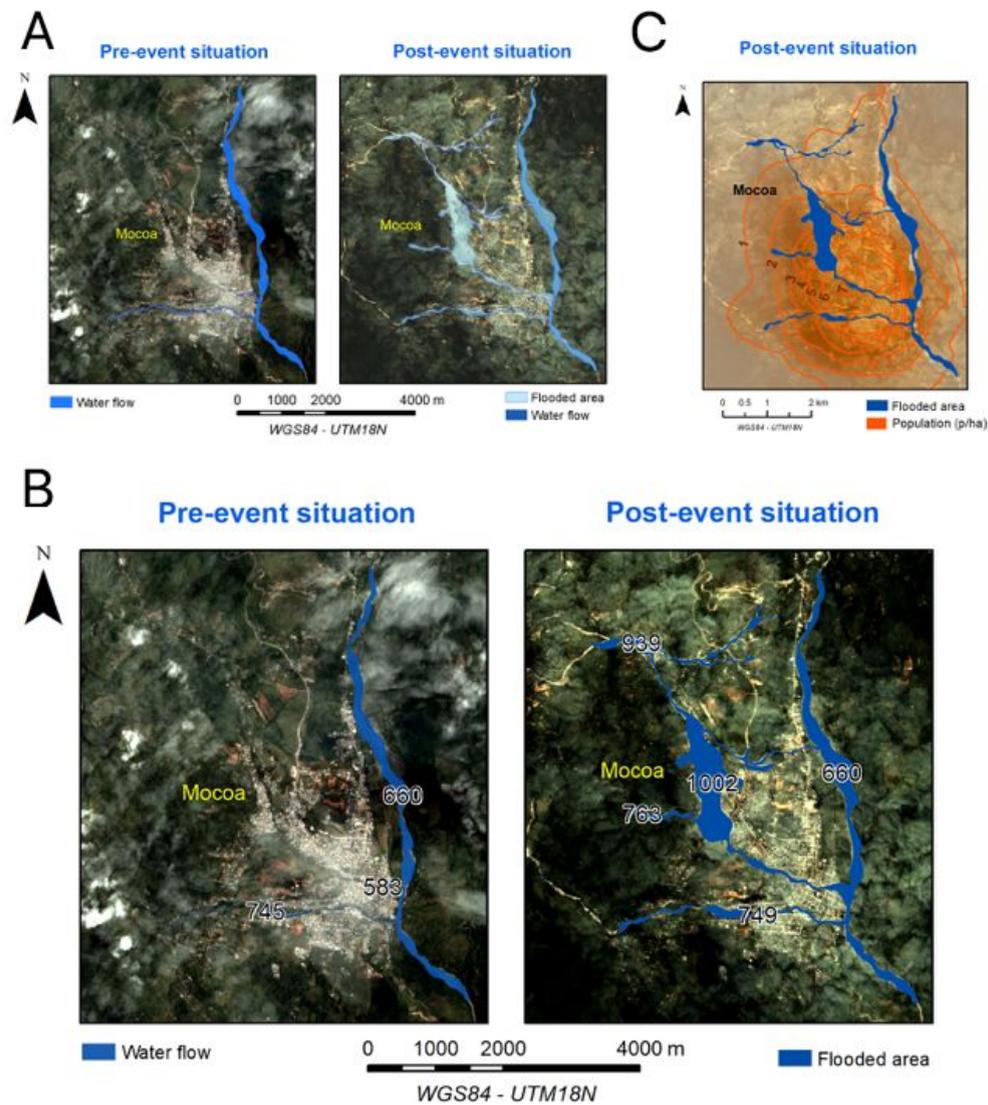

*Figure 5: A) Segmentation of the flood (right) compared to a baseline segmentation of the river Mocoa (left). B) Segmentation and elevation of the flood in both situations C) Segmented flood overlaid with a population density map extracted from worldpop.org*

**Multi-scale distribution of population during floods**

Daily aggregates of presence data were used to obtain a baseline of dynamic census for the flood in Montpellier. In order to remove weekly periodic patterns, the weekly aggregates were used to observe large-scale variations in the populations due to seasonal behavior and also potential reallocation produced by the floods as shown in Fig. 6A. Once the floods in Montpellier were detected, higher granularity (hourly) presence data was used for monitoring the movements of affected population during the disasters.

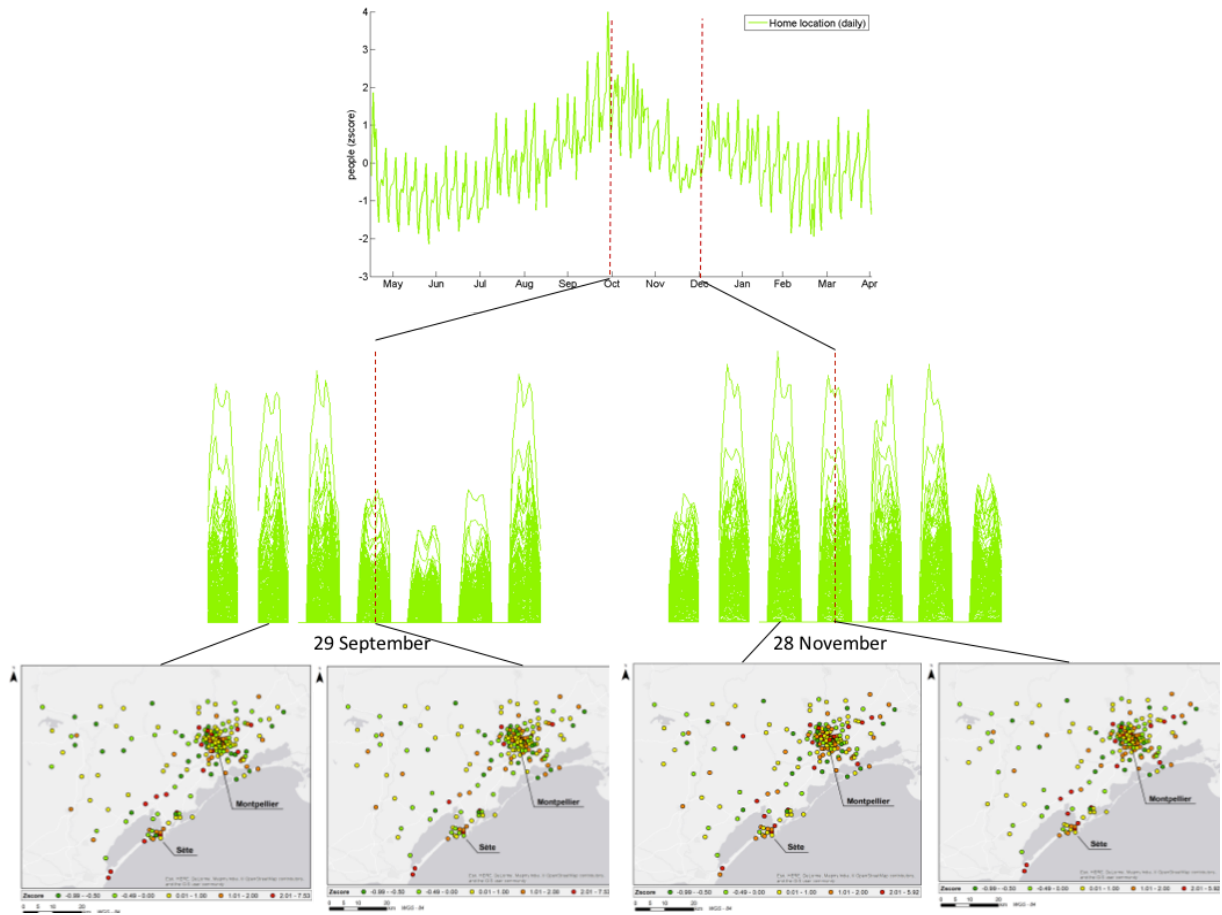

*Figure 6: Top: daily resolution aggregated presence data in Montpellier during 2014-2015. Middle: hourly resolution aggregated around the days of the floods. Bottom: Map of the z-score for two days before the flood and the day of the flood in both cases.*

We observed a very different behaviour during the first and the second flood in Montpellier that may be influenced by the large scale modulation in the population and also by the short lapse between the two floods. During the first flood we observed that the population concentration decreased as opposite to the behavior observed in other flood studies [8]. The socio-economic level of the region and the transportation infrastructure level seemed to be determining factors for the underlying mobility during a disaster. Ground transportation collapse could easily happen in developing countries whereas developed transportation infrastructure would allow a fast evacuation of the region. Fast mobility to regions away from the disaster could only be measured using trajectories crossing the affected area [21, 22].

During the second flood we observed a slightly increasing concentration of people around the day of the flood and for few antennas. Due to the difficulties to obtain a physical impact map of flood in large urban areas, it was not possible to make hypothesis about the causes of this different behavior. The socio-economic profile of population staying in the region could be an important factor for this difference.

**Socio-economic impact proxies**

Social media may also provide proxies of socio-economics during disasters. We assessed how social media could provide actionable insights in the context of the flood cases studied. The damage proxy (Methods) was used to observe how many people were concerned about insurance or damage of properties because of the floods. As shown in Figure 7, there is a very large variability described by the damage proxy. Montpellier floods caused a slight increase in the proxy 10-15 days after the floods. The damage proxy during Mocoa flood has a noisy behavior with some peaks which are apparently random. The Houston flood shows a much more clear peak of the damage proxy nearly synchronized with the awareness proxy. These results suggested that the damage of the flood as perceived by the population depends on many factors. Compared to the awareness proxy, probably triggered by immediate fear of lives, the concern about property damage and its timing largely depends on which country the flood happens.

Sentiment analysis provided by CH tools was considered, but results were highly ambiguous. During floods and for the posts filtered to compute the proxies, there was not a clear negative sentiment of the post analyzed. Although, hashtags are a popular way to broadcast trends in large-scale, natural language analysis may not able to obtain sufficient contextual information from social media posts during disasters.

For the Montpellier floods, we performed a network analysis to see the dynamics of the links created between the social media users. We gathered 8128 valid posts from the 11,123 available at CH during the period considered. A total number of 1922 users posted, from which only 545 were from female users.

Figure 8 showed the profiling of posting users and mentioned and retweeted users, disaggregated also by gender. A network model differentiating between posting users and mentioned or retweeted users (Methods, Fig. 8A) was used to analyze the linking dynamics in the month after the flood (Fig. 8B). We observed that the profiles of both types of nodes are very similar (Fig. 8C left and right) indicating some level of collective network response where the increasing activity is distributed in specific user groups. Narrowing down to the gender disaggregation, the interconnections created by the female users and male users showed some variability. While links by male users were more concentrated as temporal peals, the links created by female users seemed to span more days.

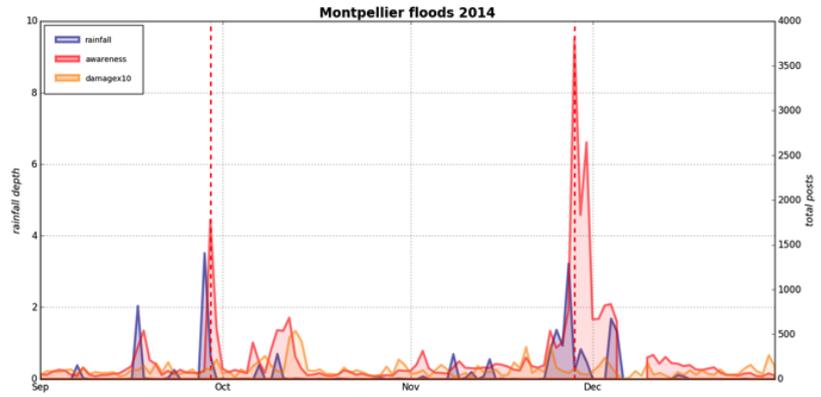
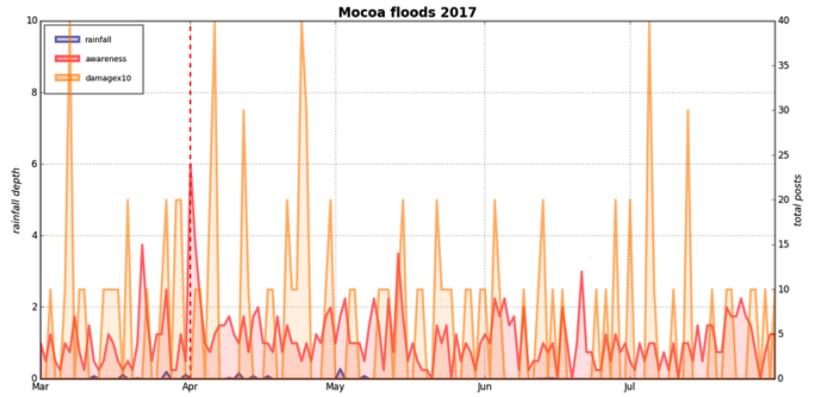
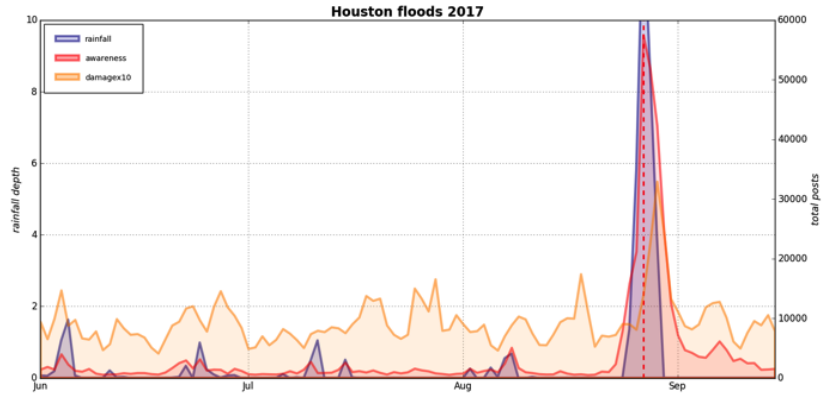

Figure 7: Floods of Montpellier, Mocoa and Houston. Blue: rainfalls measurement Red: awareness proxy Yellow: damage proxy

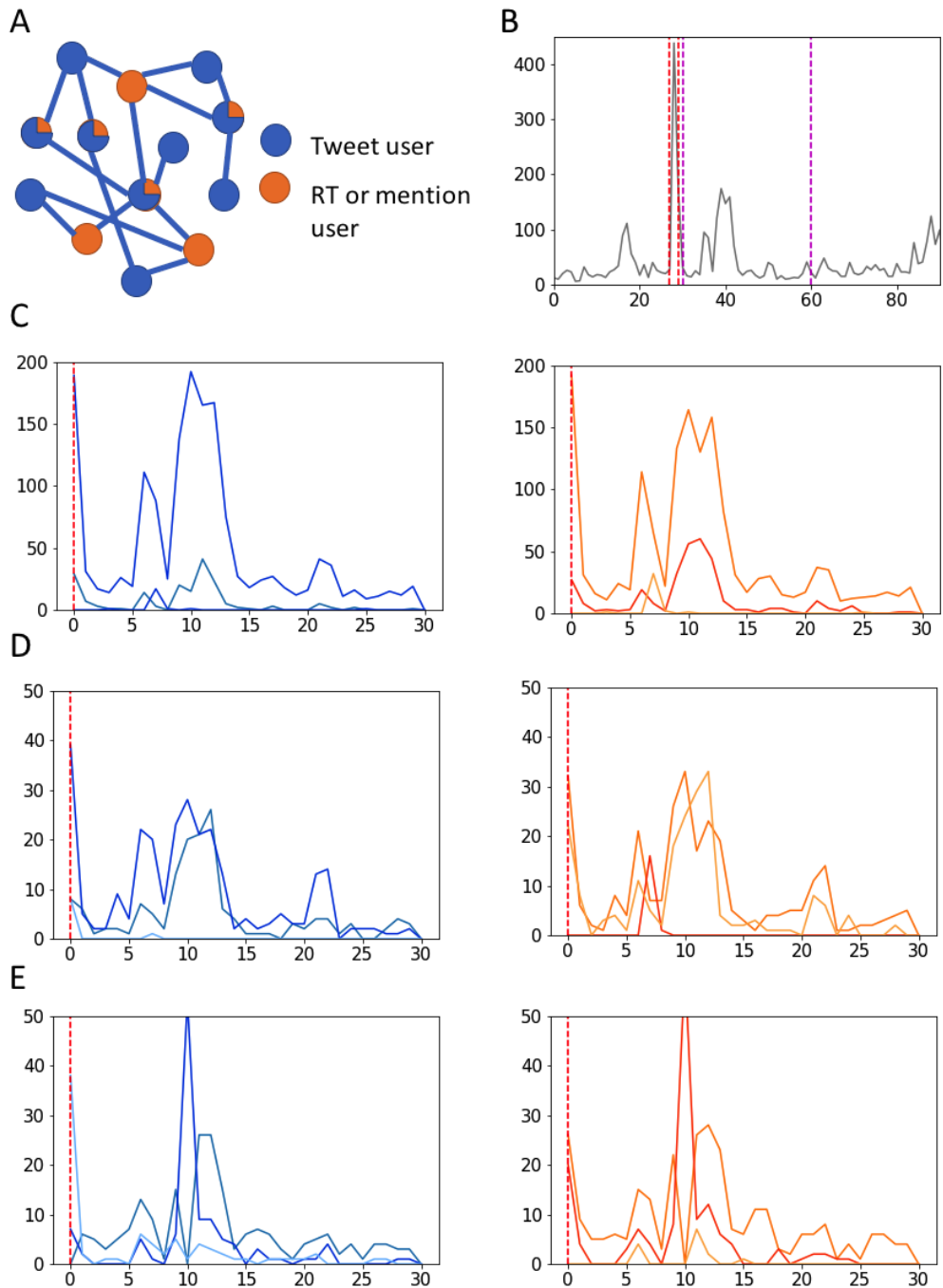

*Figure 8: A) Network model for interactions analyzed from posts B) Time series of interactions between all the nodes. Time interval for profiling is marked with vertical lines in magenta C) left: Profiling of all posting users right: Profiling of mentioned and retweeted users D) left: Profiling of all posting female users right: Profiling of mentioned and retweeted female users E) left: Profiling of all posting male users right: Profiling of mentioned and retweeted male users*

## DISCUSSION

We explored and assessed the suitability and applicability of big data sources to detect and measure the impact of different types of floods in a global scope. The framework proposed underlies the design of how to access data in real-time provided data-driven evidences for dynamic requests of high-granularity data. Social media proxies and real-time data aggregates provided by the private sector could trigger and support real-time detailed monitoring of affected areas. This could have implications for humanitarian mechanisms to be connected to different big data sources; hence, it is important to design and develop mechanisms to access data properly by keeping a rightful privacy-utility balance [23].

This work showed that for flood management, social media in combination with real-time presence data aggregates of Call Detail Records could complement environment warning mechanisms and provide an accurate proxy of awareness that matches with the maximum of rainfalls but also can serve as a detector of floods by overflow where no computational detector is available. Further analysis of social media posts was performed to provide real-time spatial, temporal, social and gender proxies suitable as evidences to trigger the request of high-granularity data to accurately estimate floods impact.

We found a severe demand of ad-hoc sensing for estimating physical impact as periodic satellite observation is not sufficient for the timely assessment of the flood dimensions. Disasters like the earthquake in Haiti triggered, for instance, the remote crowdsourcing of geo resources mapping or the release of very high resolution data such as Google eye or LiDAR. Reducing the temporal gap to image the disaster is critical, as well as evolving the detection of water bodies in adverse environmental conditions or in complex geographical locations such as cities.

Socio-economic impact proxies using social media and data aggregates were also explored. Damage proxy has been proposed before to estimate the impact of disasters in properties, however, its applicability depends on the country where the disaster occurs. Sentiment analysis is specially limited by the shortness of posts. However, social response seen from network dynamics could shed light into how a disaster affects different population groups and how social media references could help developing resilience.

Deeper assessment and evaluation of the disaster, its impact and the effectiveness of humanitarian response require analysis of exhaustive data, in most cases held by private sector, since social media activity rapidly decayed in comparison with patterns observed in CDRs. Longer term observation and innovation for resilience is prompting for data sharing mechanisms. In addition, the large variability observed encourages specific geo-contextual protocols and more social and analytical research in order to find systematic global approaches. The availability of timely data and socio-economic contextual factors are very important, but also the intrinsic behavioral dynamics prompt far more study. For instance, the initial results of this work suggested that the recursiveness or the periodicity of disasters in the same region can largely affect the behaviors and how they can be monitored to support humanitarian action.

**ACKNOWLEDGEMENTS**

We would like to acknowledge the UN Data For Climate Action Challenge for organizing the Challenge and gather the data partners. Also special thanks to United Nations Global Pulse for organizing the Challenge and being advisors of the project, specially to their Chief scientist Miguel Luengo-Oroz. We would also thank to the Challenge partners Western

Digital and Skoll Global Threats Fund. Special thanks to the data partners: Orange, Schneider Electrics, Earth Network, Crimson Hexagon and Planet. We would also thank to Africa Risk Capacity and its members Federica Carfagna and Elke Verbeeten for critical reading and fruitful collaboration. Thanks to Juan Carlos del Álamo from University California San Diego for discussion on dynamics analysis.